\documentclass{moriond}

\def\Journal#1#2#3#4{{#1} {\bf #2}, #3 (#4)}

\def\be{\begin{equation}}
\def\ee{\end{equation}}
\def\bea{\begin{eqnarray}}
\def\eea{\end{eqnarray}}

\newcommand{\angstrom}{\mbox{\normalfont\AA}}
\usepackage{siunitx}

\newcommand{\Photo}{}

\begin{document}
\vspace*{4cm}
\title{Measurement of telescope transmission using a Collimated Beam Projector}

\author{T. Souverin$^1$, J. Neveu$^{1, 4}$, M. Betoule$^1$, S. Bongard$^1$, S. Brownsberger$^2$, J. Cohen Tanugi$^{3, 7}$, S. Dagoret-Campagne$^4$, F. Feinstein$^5$, C. Juramy$^1$, L. Le Guillou$^1$, A. Le Van Suu$^6$, P. E. Blanc$^6$, F. Hazenberg$^1$, E. Nuss$^3$, B. Plez$^3$, E. Sepulveda$^1$, K. Sommer$^3$, C. Stubbs$^2$, N. Regnault$^1$, E. Urbach$^2$}

\address{$^{1}$Sorbonne Universit\'e, CNRS, Universit\'e de Paris, LPNHE, 75252 Paris Cedex 05, France; 
$^2$Department of Physics, Cambridge, Harvard University, MA 02138, USA; 
$^3$LUPM, Université Montpellier \& CNRS, F-34095 Montpellier, France; 
$^{4}$Universit\'e Paris-Saclay, CNRS, IJCLab, 91405, Orsay, France; 
$^5$CPPM, Université d'Aix-Marseille \& CNRS, 163 av. de Luminy 13288 Marseille Cedex 09, France; 
$^6$Observatoire de Haute-Provence, Université d'Aix-Marseille \& CNRS, 04870 Saint Michel L'Observatoire, France;
$^7$LPC, IN2P3/CNRS, Université Clermont Auvergne, F-63000 Clermont-Ferrand, France}

%¹LPNHE, CNRS-IN2P3 and Universités Paris 6 & 7,  ³LUPM, Université Montpellier & CNRS}

\maketitle
\abstracts{
The number of type Ia supernova observations will see a signiﬁcant growth within the next decade, especially thanks to the Legacy Survey of Space and Time undertaken by the Vera Rubin Observatory in Chile. With this rise, the statistical uncertainties will decrease and the ﬂux calibration will become the main uncertainty for the characterization of dark energy The uncertainty over the telescope transmission is a major systematic when measuring SNe Ia colors. Here we introduce the Collimated Beam Projector (CBP), a device that can measure the transmission of a telescope and its ﬁlters. Composed of a tunable monochromatic light source and optics to provide a parallel output beam this device  is able to measure with high precision the ﬁlter transmissions. In the following, we will show how measuring precisely a telescope transmission can also improve the precision of the dark energy parameters. As an example, we will present the ﬁrst results of the CBP in the context of the StarDice experiment.
}
%produce a monochromatic, homogeneous and pointlike lightsource 
\section{Collimated Beam Projector}

Type Ia supernovae (SNe Ia) are standard candles, a class of objects with predictable luminosity. By measuring the luminosity distance of SNe Ia at different redshifts, we can infer dark energy properties\cite{betoule}. We obtain this distance by measuring the maximum amplitude of the SN Ia light curve, which is observed within different restfram telescope filters depending on its redshift. The knowledge of the relative transmission of the filters is thus necessary to account for the redshift effect on SNe Ia colors and therefore constrain the dark energy parameters. 

\begin{figure}[h]
    \centering
    \includegraphics[width = 0.75\textwidth]{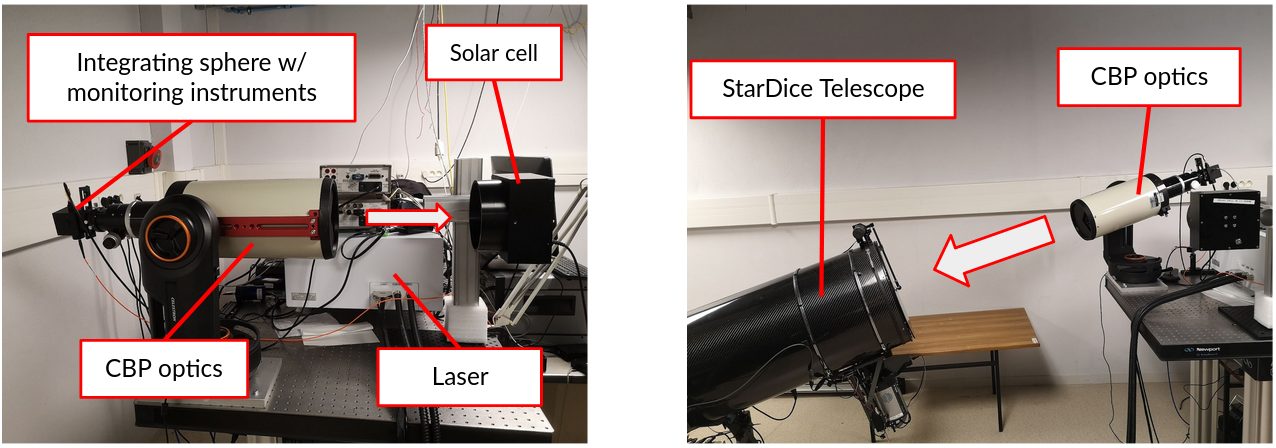}
    \caption{Left : picture of the StarDice Telescope at LPNHE (Paris). Right : picture of the CBP device, pointing toward the StarDice telescope.}
    \label{fig:CBP}
\end{figure}

\noindent The Collimated Beam Projector (CBP) is a device that measures the transmission of a telescope and its ﬁlters. It is composed of a tunable monochromatic laser (Ekspla NT252), that emits light within a range of [350 - 1100] nm, with a resolution of \SI{1}{\nano\meter} and an accuracy of \SI{1}{\angstrom}. The light is injected with an optical fiber into an integrating sphere, whose output is a pinhole of variable size, producing a monochromatic, homogeneous and pointlike lightsource. The pinhole is set at the focal point of a 152mm Ritchey-Chr\'etien\footnote{Ritchey-Chr\'etien Omegon Pro RC 154/1370} mounted backwards, thus providing a parallel beam. A photodiode and a spectrograph monitor the surface brightness and wavelength of the light inside the sphere. The response of the CBP optical device $R_\mathrm{CBP}(\lambda)$ is measured by shooting directly into a flux calibrated solar cell\cite{brownsberger}, as the ratio of the charges detected in the solar cell over the charges detected  the monitoring photodiode (fig.\ref{fig:cbp_plot}).

\section{Telescope calibration}

\begin{figure}[h]
    \centering
    \includegraphics[width = 1\textwidth]{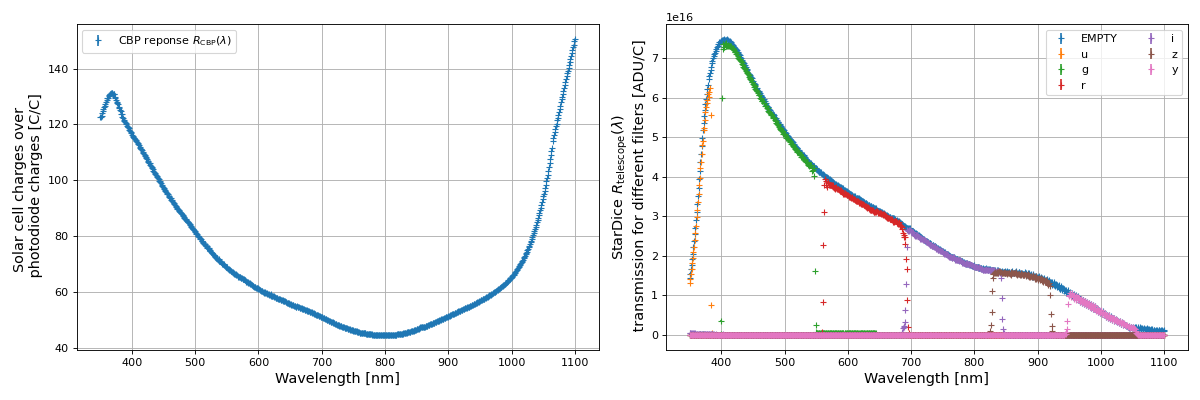}
    \caption{Left: measurement of the response of the CBP optical device in a range of [350-1100] nm. Right:  preliminary measurement of ﬁlter transmissions of the StarDice telescope obtained with the CBP.  }
    \label{fig:cbp_plot}
\end{figure}

The goal is to measure the StarDice\cite{hazenberg} telescope throughput $R_{\mathrm{telescope}}(\lambda)$, which is a product of the telescope mirror reflectivities, the filter optical transmissions and the quantum efficiency of the CCD camera. To calibrate a telescope with the CBP, we shoot the parallel light beam into this instrument, and we measure the detected flux $\phi_{\mathrm{obs}}$ on the CCD camera doing aperture photometry and dark subtraction. On the other hand we can monitor the emitted flux $\phi_{\mathrm{source}}$ thanks to the CBP photodiode. The relation between these two values is:
\begin{equation}
    \phi_{\mathrm{obs}} = \phi_{\mathrm{source}} \times R_{\mathrm{CBP}}(\lambda) \times R_{\mathrm{telescope}}(\lambda)
    \label{eq:cbp}
\end{equation}
\noindent This measurement have been made at LPNHE, our best result so far is shown in figure \ref{fig:cbp_plot}. Systematic uncertainties are under investigation, but the precision should be around a few per mil.

\noindent The LSST telescope will increase significantly the number of SNe Ia observed. François Hazenberg\cite{hazenberg} has shown in his thesis that this new dataset can improve the dark energy parameter constraint by a factor of 10, only if we know the filter transmissions of the LSST telescope with a precision greater than 0.1\%. Thanks to the CBP, we demonstrated that we can reach such a precision on the StarDice filters, and our goal is to do likewise with LSST.

\end{document}